\documentclass[aip,pop,reprint]{revtex4-1}
\usepackage{graphicx}
\usepackage{ulem}
\usepackage{amssymb}
\usepackage{amsmath}
\usepackage{todonotes}
\usepackage{wasysym}
\usepackage{relsize}
\usepackage{bm}
\usepackage{textcomp}
\usepackage{adjustbox}

\begin{document}

\title{Simulating multiscale gated field emitters - a hybrid approach}

\author {Shreya Sarkar}
\affiliation {Homi Bhabha National Institute, Mumbai 400 094, INDIA}
\affiliation {Bhabha Atomic Research Centre,Mumbai 400 085, INDIA}
\author{Raghwendra Kumar}
\affiliation {Bhabha Atomic Research Centre,Mumbai 400 085, INDIA}
\author{Gaurav Singh}
\affiliation {Homi Bhabha National Institute, Mumbai 400 094, INDIA}
\affiliation {Bhabha Atomic Research Centre,Mumbai 400 085, INDIA}
\author{Debabrata Biswas}\email{dbiswas@barc.gov.in}
\affiliation {Homi Bhabha National Institute, Mumbai 400 094, INDIA}
\affiliation {Bhabha Atomic Research Centre,Mumbai 400 085, INDIA}

\begin{abstract}
  
  Multi-stage cathodes are promising candidates for field emission due to the multiplicative effect in local field
  predicted by the Schottky conjecture and its recent corrected counterpart [J. Vac. Sci. Technol. B 38, 023208 (2020)].
Due to the large variation in length scales even in a 2-stage compound structure consisting of a macroscopic
base and a microscopic protrusion, the simulation methodology of a gated field emitting compound diode needs to be
revisited. As part of this strategy, the authors investigate the variation of local field on the surface of a
compound emitter near its apex and find that the generalized cosine law continues to hold locally near the tip
of a multi-scale gated cathode. This is used to emit electrons with appropriate distributions in
position and velocity components  with a knowledge of only the electric field at
the apex. The distributions are consistent with contemporary free-electron field emission model
and follow from the joint distribution of launch angle, total energy, and normal energy.
For a compound geometry with local field enhancement by a factor of around 1000, 
a hybrid model is used where the vacuum field calculated using COMSOL is imported into the
Particle-In-Cell code PASUPAT where the emission module is implemented. Space charge effects
are incorporated in a multi-scale adaptation of PASUPAT using a truncated geometry with
`open electrostatic boundary' condition. The space charge field, combined with the vacuum field, is used
for particle-emission and tracking.

\end{abstract}

\maketitle
\newcommand{\be}{\begin{equation}}
\newcommand{\ee}{\end{equation}}
\newcommand{\bea}{\begin{eqnarray}}
\newcommand{\eea}{\end{eqnarray}}
\newcommand{\Tbar}{{\bar{T}}}
\newcommand{\En}{{\cal E}}
\newcommand{\K}{{\cal K}}
\newcommand{\GC}{{\cal \tt G}}
\newcommand{\Lop}{{\cal L}}
\newcommand{\DB}[1]{\marginpar{\footnotesize DB: #1}}
\newcommand{\q}{\vec{q}}
\newcommand{\kt}{\tilde{k}}
\newcommand{\Lopn}{\tilde{\Lop}}
\newcommand{\noi}{\noindent}
\newcommand{\ovn}{\bar{n}}
\newcommand{\ovx}{\bar{x}}
\newcommand{\ovE}{\bar{E}}
\newcommand{\ovV}{\bar{V}}
\newcommand{\ovU}{\bar{U}}
\newcommand{\ovJ}{\bar{J}}
\newcommand{\calE}{{\cal E}}
\newcommand{\ovphi}{\bar{\phi}}
\newcommand{\zt}{\tilde{z}}
\newcommand{\rt}{\tilde{\rho}}
\newcommand{\tth}{\tilde{\theta}}
\newcommand{\nuv}{{\rm v}}
\newcommand{\ck}{{\cal K}}
\newcommand{\cc}{{\cal C}}
\newcommand{\ca}{{\cal A}}
\newcommand{\cb}{{\cal B}}
\newcommand{\cg}{{\cal G}}
\newcommand{\ce}{{\cal E}}
\newcommand{\csa}{{\cal a}}
\newcommand{\fn}{{\small {\rm  FN}}}
\newcommand\norm[1]{\left\lVert#1\right\rVert}

\section{Introduction}
\label{sec:intro}

The revival of interest in the century-old Schottky
Conjecture (SC) \cite{Schottky23,stern,miller09,deAssis16,marcelino,harris19a,zhu2019,marcelino19,DB_CSC_20},
and its recent
experimental demonstration using the field emission data from a suitably constructed
compound cathode\cite{CT_2020}, point towards the potential utility of the multiplicative
effect in constructing electron guns.
The Schottky conjecture essentially states that if a protrusion exists on top
of a large curved base, the electrostatic field enhancement factor at the tip of the protrusion
is the product of the individual apex enhancement factors of the base ($\gamma_a^{(1)}$)
and the protrusion ($\gamma_a^{(2)}$), provided the protrusion is sufficiently small.
Its utility has been recently enhanced
by the Corrected Schottky Conjecture (CSC) which accounts for situations when the
protrusion is not very small and states that the apex enhancement factor at the tip
of the compound structure, $\gamma_a^{(C)} = \gamma_a^{(2)} \langle \gamma_a^{(1)} \rangle$
where $\langle \ldots \rangle$ denotes the average enhancement over the height of the protrusion. 
In the limiting case when the protrusion is infinitesimally small $\langle \gamma_a^{(1)} \rangle = \gamma _a^{(1)}$
so that $\gamma_a^{(C)} = \gamma_a^{(2)}  \gamma_a^{(1)} $, while as the height of the protrusion becomes large
$\langle \gamma_a^{(1)} \rangle << \gamma_a^{(1)}$ and the effect of multiplication tends to vanish.
Both the SC\cite{Schottky23} and the CSC\cite{DB_CSC_20} deal with situations when the anode is placed far away.
It has recently been
verified\cite{CT_2020} that the CSC in fact holds even when the anode is in close proximity.

In order to make use of the multi-stage cathode in constructing devices, it is important
that the multiplicative effect continues to hold in gated structures. Assuming that it does, the enhanced local electrostatic
field can aid field emission by lowering the turn-on fields and generating high current densities
at moderate applied macroscopic fields. Clearly, there is a need to study all aspects of field emission from
gated compound structures.

The evaluation of the net field emission current requires us to know the local field normal to the
surface of the emitter in the vicinity of the apex. While this can be easily determined using
a finite element code such as COMSOL, most Particle-In-Cell (PIC) codes use finite differencing algorithms
due to the ease in implementing Maxwell's equations consistently. A high aspect ratio field emitter
thus requires a lot of resources in dealing with the emission process since local fields on the
curved surface near the emitter apex must be determined accurately\cite{jensen2019a}.
A way around  this difficulty is to find only the field at the apex accurately
and use analytical results for smooth parabolic tips to emit electrons or macro-particles
having the correct distribution in position and velocity. 
This approach relies on the `universal' generalized cosine law of local field variation
for locally parabolic tips. It states that for an axially symmetric emitter described by
$z = z(\rho)$, the electric field on the surface near the apex\cite{ultram,physE}

\be
E(\rho) = E_a \frac{z/h}{\sqrt{(z/h)^2 + (\rho/R_a)^2}} = E_a \cos\tth   \label{eq:cos}
\ee
  
\noi
where $E_a$ is the field at the apex, $h$ is the height of the emitter and $R_a$ is
the radius of curvature at the apex. Thus, if the field at the apex is known, the cosine
law can be used to determine the net field emission current ($I_{MG}$) within
the Fowler-Nordheim\cite{FN28} and Murphy-Good\cite{MG56} formalisms\cite{RF2006,RFD2007,KLJ2014,DBRR2019,RF2019}.
Importantly, for purposes of simulation, the cosine law
can also be used to determine the current distributions\cite{DB_parabolic,egorov17} $f(\tth)$, $f(\calE_N|\tth)$ and
$f(\calE_T|\tth,\calE_N)$ from the joint distribution $f(\tth,\calE_N,\calE_T)$
where $\calE_N$ and $\calE_T$ are the normal and total energies of the electrons.
These distributions can thus be used to launch particles with
the distributions in $\rho$ (equivalently $\tth$ or $z$) and the velocity components $v_x,v_y$ and $v_z$.
The emission process can thus be handled without using up precious resources required
for simulating the rest of the domain.  
Field emission from gated multi-stage compound cathodes can similarly be simulated
if the generalized cosine law  is found to hold.

\begin{figure}[ht]
  \begin{center}
   \vskip -1.85cm
\hspace*{-.30cm}\includegraphics[scale=0.4,angle=0]{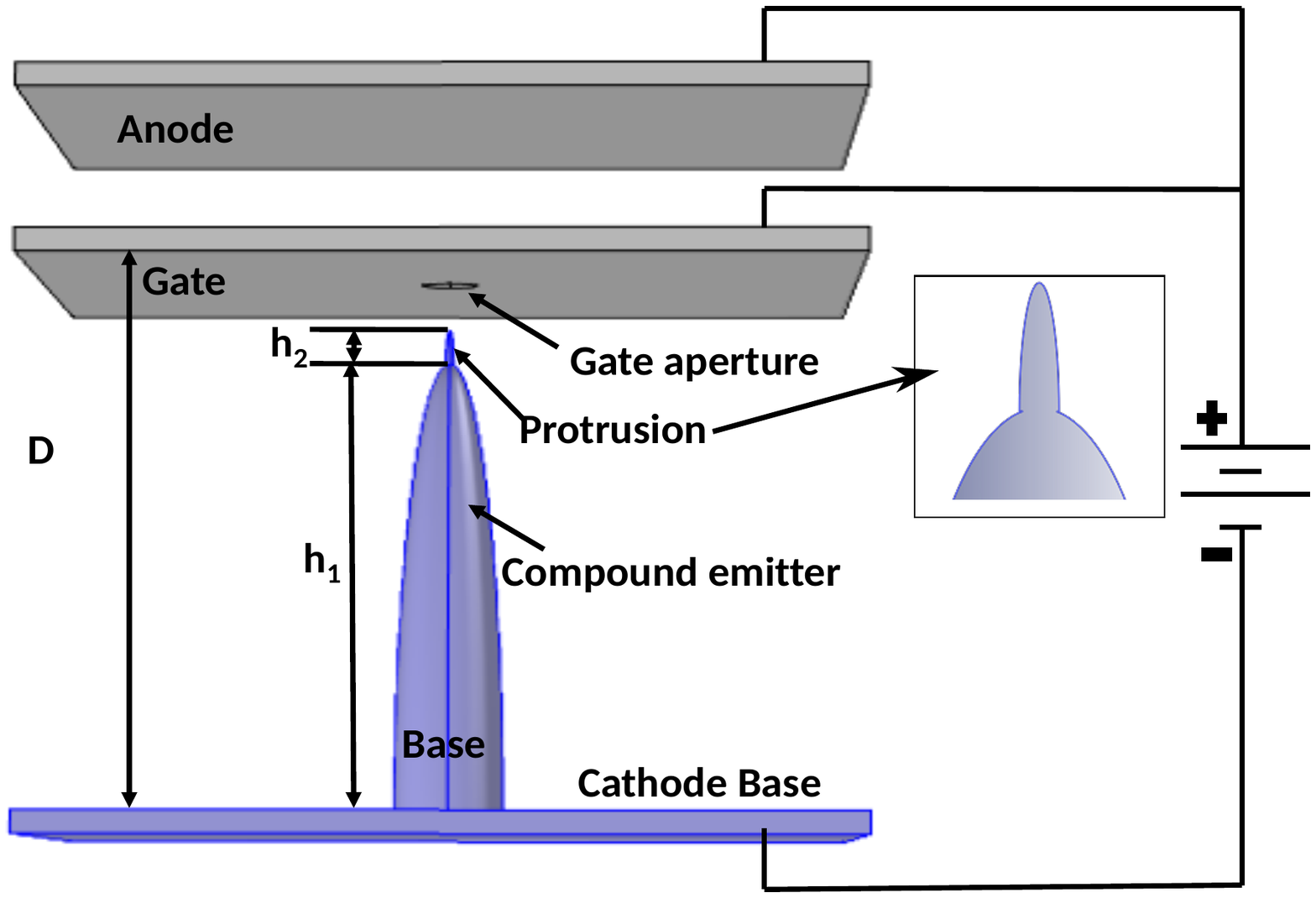}
\vskip -5.0cm
\caption{A schematic of the gated triode with a compound multi-stage emitter consisting of
  a base and a protrusion. The anode and the gate are kept at the same potential for
  the calculations presented here. 
  }
\label{fig:schematic}
  \end{center}
\end{figure}

The plan of the paper is as follows. We shall first confirm the validity of the
multiplicative effect in gated multi-stage compound cathodes (such as in Fig.~\ref{fig:schematic})
and verify how well
it is described by the SC and CSC as the gate-radius ($R_g$) and gate-position ($D$) are changed.
We shall then study the validity of the generalized cosine law under various
gate conditions and protrusion height. Thereafter, we shall use the
current distributions that follow on using the generalized cosine law along with the relevant
field emission equation for current density, to emit a beam of electrons from
the tip of the compound cathode having a distribution of velocity components.
A hybrid approach is used in the simulation wherein the vacuum electrostatic field
is imported from COMSOL to the PIC code PASUPAT which has an implementation of
the emission module discussed above. Space-charge effects are incorporated by using a truncated geometry
where the macroscopic base and cathode plate are removed and replaced by
open electrostatic boundary condition. The vacuum field is then added to the space
charge field to emit and track particles self-consistently.

\section{Multiplicative effects in a gated device}
\label{sec:csc}

To the best of our knowledge, multiplicative effects in the local electric field on the
surface of a multi-scale gated structure have not been analyzed before. We shall consider a 2-stage
compound structure in the following and explore the validity of the Schottky conjecture and its recent corrected
counterpart which is appropriate when the protrusion is not negligible in height.

The importance of a compound cathode stems from the fact that a high local field enhancement
requires the ratio $h/R_a$ to be large. For an isolated emitter, with the anode
and other emitters far away from its neighborhood, the apex field enhancement
factor (AFEF) $\gamma_a = E_a/E_0$ is expressed as \cite{DB_universal}

\be
\gamma_a = \frac{2h/R_a}{\alpha_1 \ln(4h/R_a) - \alpha_2} \label{eq:gam0}
\ee

\noi
where $\alpha_{1,2}$ are constants for a particular geometry\cite{hemiellip}, $h$ is the height of the
emitter and $R_a$ is its apex radius of curvature. The local field at the apex
$E_a = \gamma_a E_0$ where $E_0$ is the applied or macroscopic field. In a parallel
plate geometry with a curved emitter mounted on the cathode plate, $E_0 = V_g/D$ where
$V_g$ is the potential difference across the diode and the $D$ is the plate separation.
Clearly, a high field enhancement requires the structure to be tall and sharp.
For a fixed apex radius of curvature $R_a$, a single tall structure is mechanically
unstable while a compound multi-stage cathode of the same height and apex radius of
curvature is more stable and can achieve the desired field enhancement. A multi-stage
construction is thus desirable.

\subsection{Compounding and local field multiplication}

For the 2-stage multiscale compound cathodes considered here, the apex field
enhancement factor $\gamma_a^{C}$ is calculated using COMSOL. For purposes of
using SC or CSC, the individual enhancement factors of the base ($\gamma_a^{(1)}$)
and the protrusion ($\gamma_a^{(2)}$) are also determined using COMSOL under identical
gate conditions including the aperture radius $R_g$. Thus, for calculating $\gamma_a^{(1)}$, the protrusion is removed
keeping the rest of the geometry unchanged, while for evaluating $\gamma_a^{(2)}$, the
protrusion is placed directly on the cathode plate and the gate is kept at a distance
$D - h_1$ from the cathode plate. 
In each case, convergence is ensured by varying
the mesh parameters in COMSOL.

\begin{figure}[hbt]
  \begin{center}
   \vskip -0.75cm
\hspace*{-0.70cm}\includegraphics[scale=0.35,angle=0]{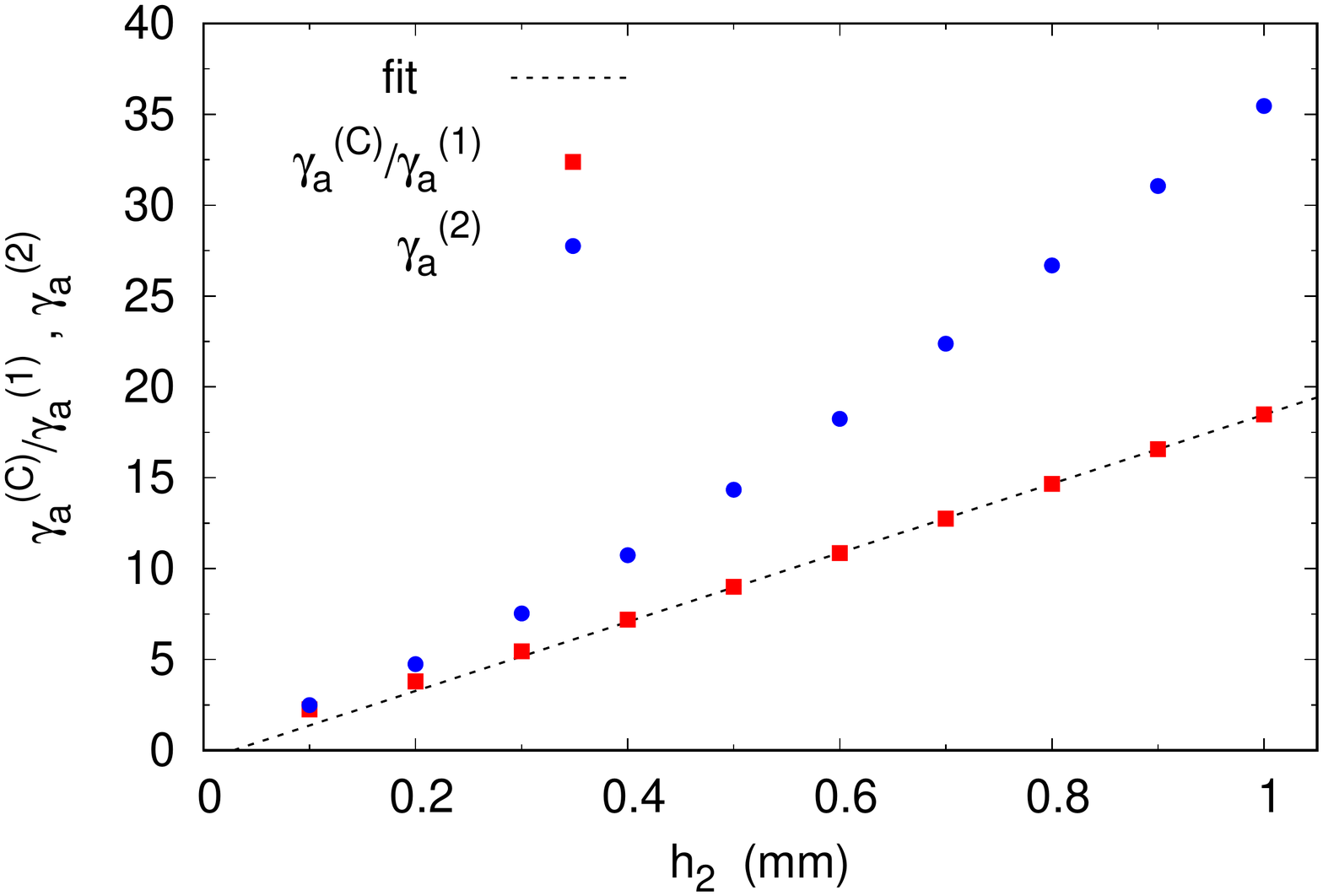}
\vskip -1.0cm
\caption{The AFEF ratio of the compound structure and base $\gamma_a^{(C)}/\gamma_a^{(1)}$ is plotted
  (solid squares) alongside the AFEF of the protrusion $\gamma_a^{(2)}$ (solid circles). When
  the protrusion height is small $\gamma_a^{(C)}/\gamma_a^{(1)}$ coincides with $\gamma_a^{(2)}$.
  For larger $h_2$, the multiplication factor appears to be linear as seen from the
  fitted line (dashed). Here $h_1 = 50$mm, $b_1 = 5$mm, $b_2 = 0.1$mm, $D = 51$mm and $R_g = 1$mm.
  }
\label{fig:gamma_vs_h2_1}
  \end{center}
\end{figure}

In order to ascertain the multiplicative enhancement of local field, consider
a curved axially symmetric primitive base having a height $h_1 = 50$mm and base radius $b_1 = 5$mm
shaped as a hemiellipsoid. It is placed 
in a gated configuration as shown in the Fig.~\ref{fig:schematic}
with the gate  placed at a distance $D = 51$mm from the cathode plane and having an aperture
of radius $R_g = 1$mm. The axis of the hemiellipsoid passes through the centre of the gate-aperture.
Since the  primitive base is an hemiellipsoid, its apex radius of curvature $R_a^{(1)} = b_1^2/h_1 = 0.5$mm.
At the next stage, we place another hemiellipsoid (protrusion or crown) of base radius $b_2 = 0.1\mu$m
which is small compared to $R_a^{(1)}$. The height of the protrusion $h_2$ is varied
from $0.1$mm to $1$mm.

\begin{figure}[hbt]
  \begin{center}
    \vskip -0.75cm
\hspace*{-0.70cm}\includegraphics[scale=0.35,angle=0]{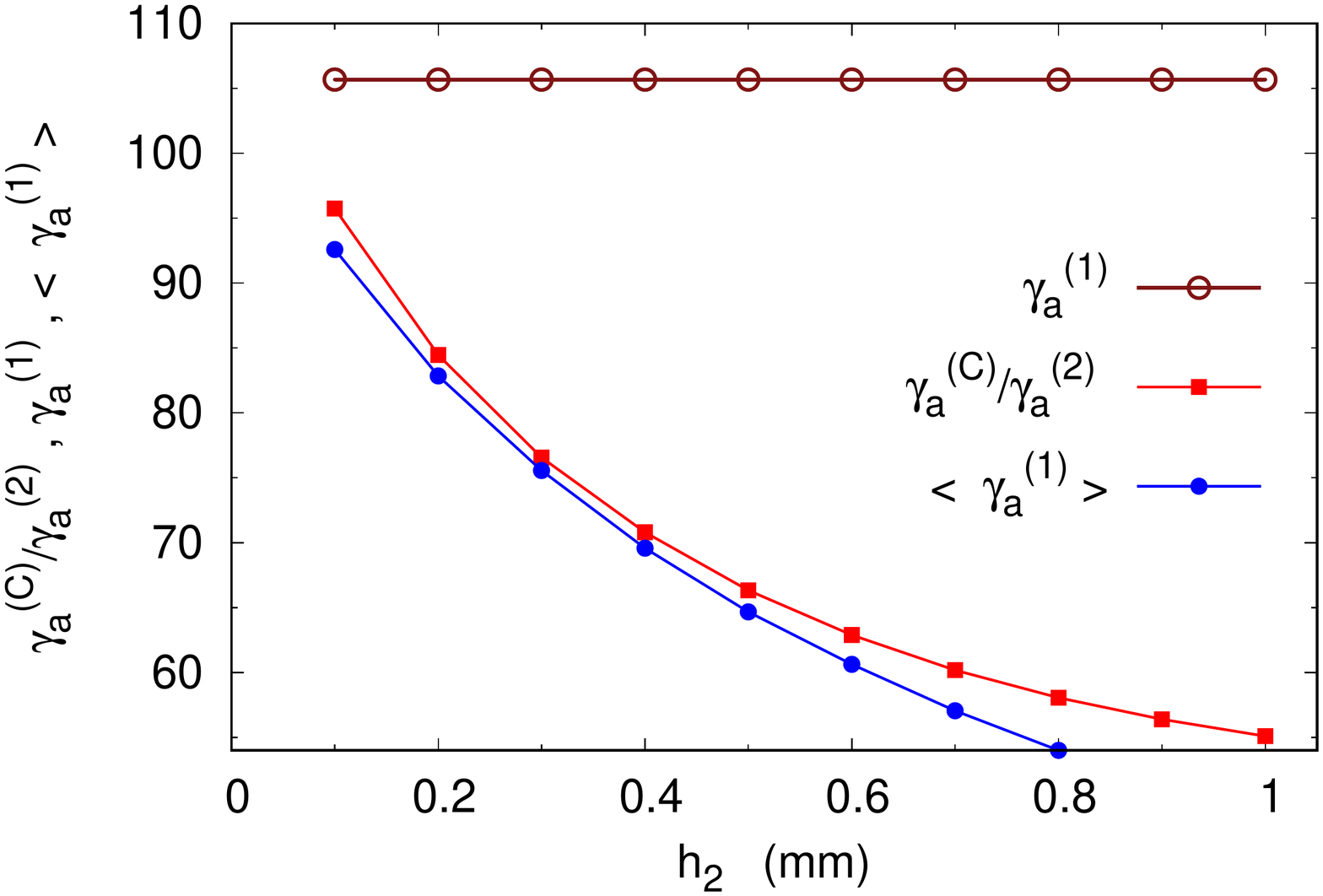}
\vskip -1.0cm
\caption{The AFEF ratio of the compound structure and protrusion $\gamma_a^{(C)}/\gamma_a^{(2)}$ (solid squares)
  is plotted alongside the AFEF of the base $\gamma_a^{(1)}$ (open circles). Also shown is
  the average enhancement of the base $\langle \gamma_a^{(1)} \rangle$ (solid circles)
  given by Eq.~(\ref{eq:av}) which is relevant for the Corrected Schottky Conjecture.
  Here $h_1 = 50$mm, $b_1 = 5$mm, $b_2 = 0.1$mm, $D = 51$mm and $R_g = 1$mm.
  }
\label{fig:gamma_vs_h2_2}
  \end{center}
\end{figure}

The variation of the apex field enhancement factor of the compound
structure normalized to that of the primitive base is shown in Fig.~\ref{fig:gamma_vs_h2_1}.
The solid squares are the values of $\gamma_a^{(C)}/\gamma_a^{(1)}$ where $\gamma_a^{(C)}$ and
$\gamma_a^{(1)}$ are the AFEF of the compound structure and the the primitive base respectively.
Also shown alongside is the AFEF of the protrusion $\gamma_a^{(2)}$ using solid circles.
Note that $\gamma_a^{(1)}$ is evaluated keeping the separation between the plates ($D$) as in the
compound structure, while $\gamma_a^{(2)}$ is evaluated keeping the distance between the gate
and cathode plate as $D - h_1$.

Clearly, a value $\gamma_a^{(C)}/\gamma_a^{(1)} > 1$ shows the multiplicative effect. Indeed, for
small $h_2$, the multiplication factor coincides with the AFEF of the protrusion on top
of the base. This is consistent with the Schottky conjecture. As $h_2$ increases, the
multiplication factor drops below the value of the AFEF of the protrusion ($\gamma_a^{(2)}$), consistent
with the expectation from the Corrected Schottky Conjecture.

The multiplicative effect can be alternately presented to show how a pre-fabricated protrusion
benefits by being placed on top of a base. We thus plot $\gamma_a^{(C)}/\gamma_a^{(2)}$ for the same
geometry in Fig.~\ref{fig:gamma_vs_h2_2}. The base is held at a fixed height $h_1 = 50$mm
while the height of the protrusion is varied. A long protrusion (e.g. $h_2/R_a^{(1)} = 2$)
also enjoys the enhancement of the base in a gated structure, albeit with a reduced
multiplicative effect. 

\subsection{Predictions of SC and CSC for gated structures}

It is thus clear that multi-stage gated cathodes can be used to enhance the local field. We shall
now briefly study the errors in prediction of $\gamma_a^{(C)}$ using the Schottky Conjecture and the
Corrected Schottky Conjecture.
Since the radius of the gate aperture changes the local field at the apex of the
compound structure, the predictive capabilities of SC and CSC
under different gate conditions also need to be investigated.

\begin{figure}[hbt]
  \begin{center}
    \vskip -0.75cm
\hspace*{-0.70cm}\includegraphics[scale=0.35,angle=0]{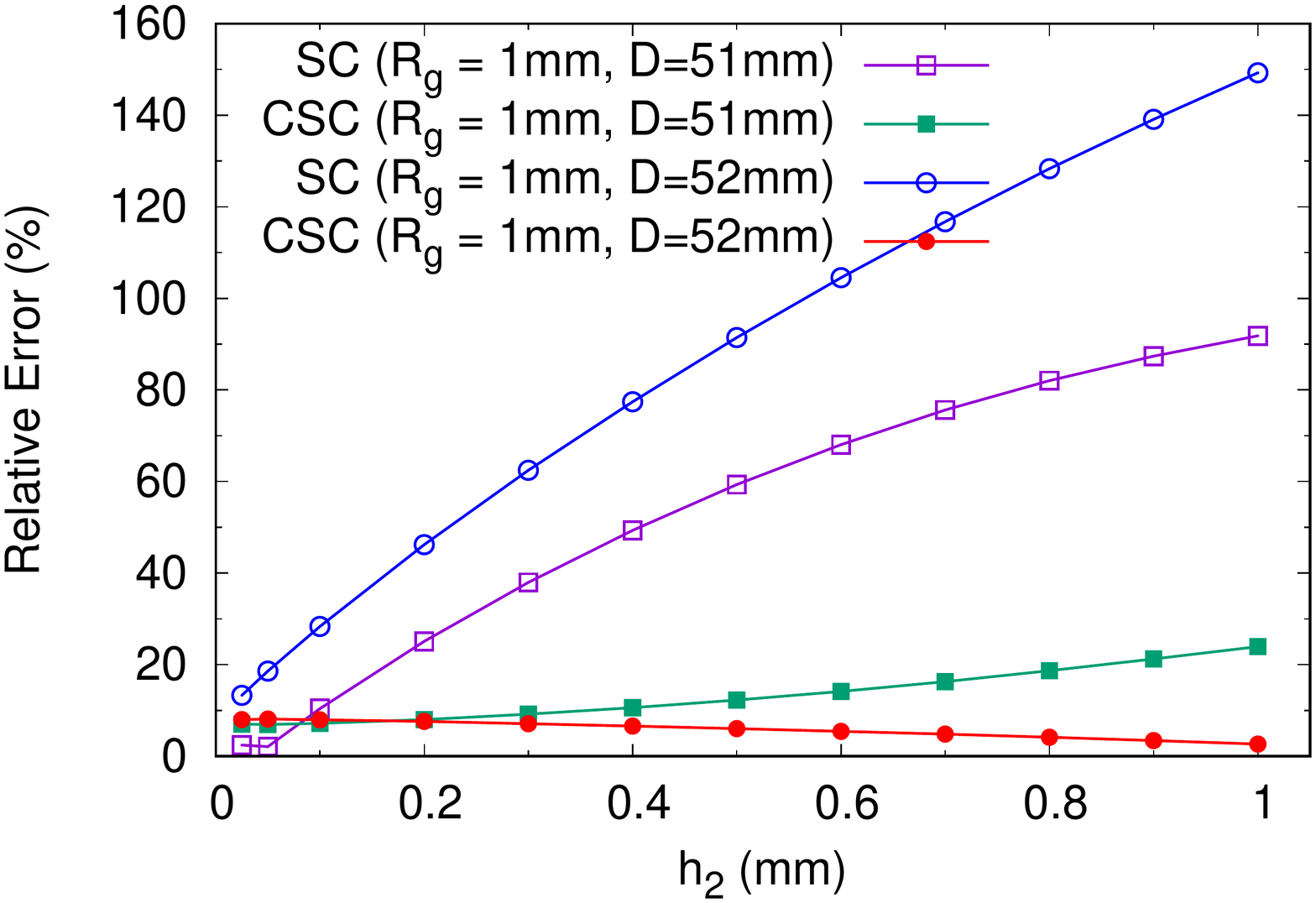}
\vskip -1.0cm
\caption{The variation in relative errors in AFEF, calculated using SC and CSC,  with the height
  of the protrusion for gate radius $R_g = 1$mm.
  This is shown for gates placed at $D = 51$mm and $D = 52$mm.
  }
\label{fig:error_Rg1mm}
  \end{center}
\end{figure}

Consider thus a primitive hemiellipsoid base of height 50mm and base radius 5mm. The protrusion
is also a hemiellipsoid with base $0.1$mm and its height $h_2$ is varied from 0.025mm to 1mm. The gate radius
is kept fixed at $R_g = 1$mm.
Fig~\ref{fig:error_Rg1mm} shows respectively the relative error in SC and CSC predictions for
gate placed at 
a distance $D = 51$mm and $D = 52$mm from the cathode plane.

The relative error for the Schottky conjecture is defined as
$100 \times |\gamma_a^{(C)} - \gamma_a^{(1)} \gamma_a^{(2)}|/\gamma_a^{(C)}$
while in case of the Corrected Schottky Conjecture, the relative error is 
$100 \times |\gamma_a^{(C)} - \langle \gamma_a^{(1)} \rangle  \gamma_a^{(2)}|/\gamma_a^{(C)}$
where

\be
\langle \gamma_a^{(1)} \rangle = \frac{1}{E_0} \int_{h_1}^{h_1 + h_2}~E_z(z)~dz  \label{eq:av}
\ee

\noi
is the average field enhancement\cite{DB_CSC_20} on the axis over the height of the protrusion $h_2$.
Clearly, for both gate positions, the CSC performs well while the SC is comparable or performs
better for very small protrusions. Note that while the gate position ($D$) and aperture radius ($R_g$)
have an effect on the relative
error in both the SC and CSC prediction, the CSC in general performs much better in all the cases
studied \cite{csc_averaging}.

We can thus conclude that the multiplicative effect  is pronounced in a multi-scale cathode
and the Corrected Schottky Conjecture can be reliably used to predict the
apex field enhancement factor.

\section{Local field variation and the cosine law}

We shall now study the variation of surface local field near the apex of the
multi-scale compound structure. As in section \ref{sec:csc}, we shall consider
a hemiellipsoid protrusion on top of a hemiellipsoid base as an example.
In addition, we shall also consider an HECP (hemiellipsoid on a cylindrical post) protrusion
on a hemiellipsoid base.

\begin{figure}[hbt]
  \begin{center}
    \vskip -4.35cm
\hspace*{-0.40cm}\includegraphics[scale=0.465,angle=0]{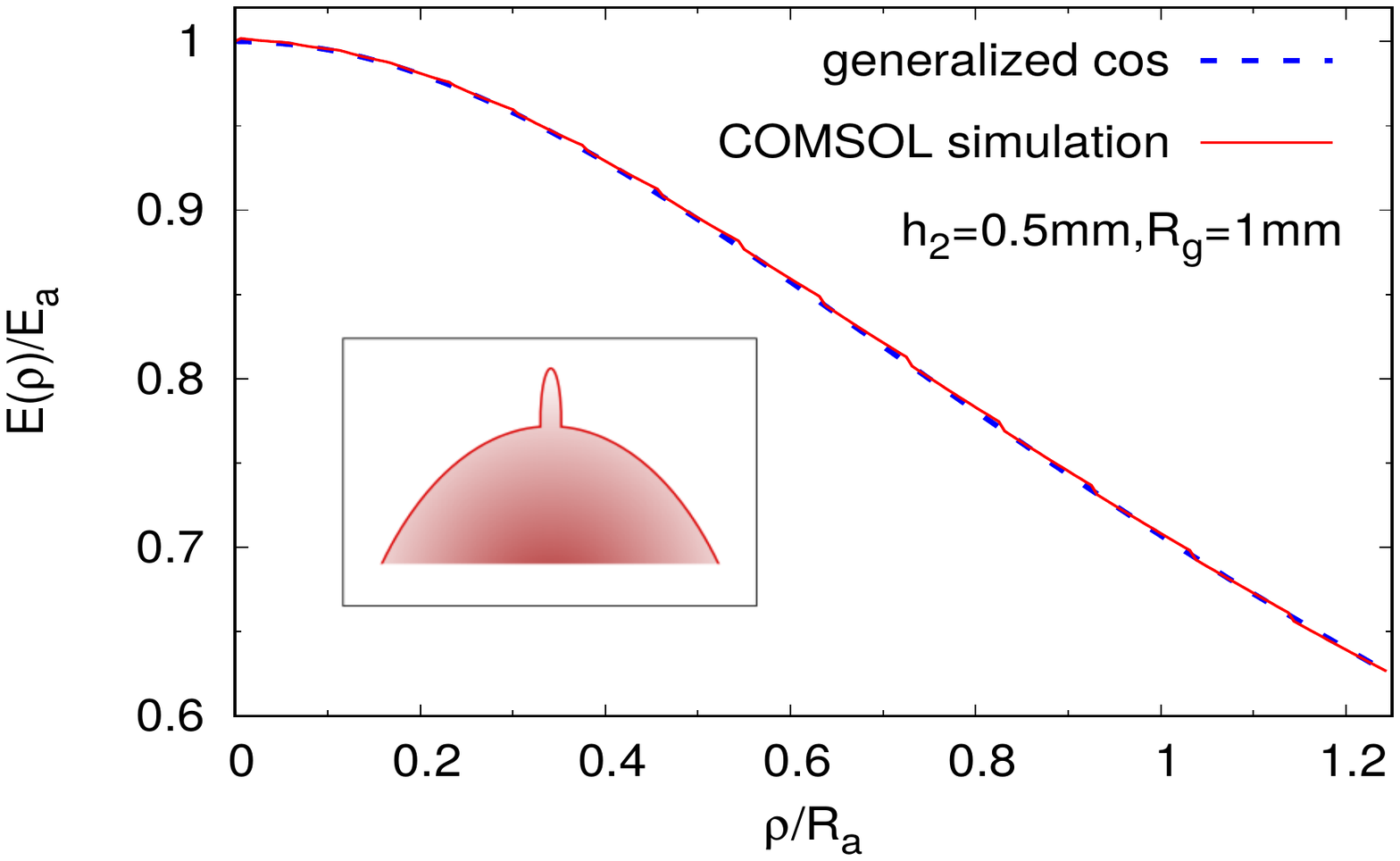}
\vskip -4.5cm
\caption{The variation in the local field normalized to the field at the apex
  of the protrusion (denoted by $E_a$) along the surface of the protrusion in
  a 2-stage compound geometry comprising of 2 hemiellipsoids. Here $R_a = R_a^{(2)}$ denotes
  the apex radius of curvature of the compound geometry. The protrusion height $h_2 = 0.5$mm
  while $D = 51$mm. Inset shows part of the ellipsoid-base and the protrusion.
  }
\label{fig:cos_hp0.5_Rg1}
  \end{center}
\end{figure}

\vskip -0.50cm
Note that the generalized cosine law of Eq.~(\ref{eq:cos}) was first derived analytically for
a hemiellipsoid with the anode far away. It was also shown to hold numerically
for other shapes \cite{ultram} and with the anode in close proximity \cite{db_anode}.
It was subsequently established analytically\cite{physE} using the nonlinear line
charge model\cite{db_jap2016,zheng2020} that the cosine law holds approximately for sharp parabolic emitters ($h/R_a >> 1$).
While a multistage cathode is expected to have a non-linear line charge density \cite{surface2line},
it may not be smooth at the junction of two stages. It is thus necessary to ascertain the validity
of the generalized cosine law of local field variation.

Fig.~\ref{fig:cos_hp0.5_Rg1} shows the variation on the surface of a hemiellipsoid protrusion
of height $h_2 = 0.5$mm, base radius $b_2 = 0.1$mm,  placed on a hemiellipsoid base of height
$h_1 = 50$mm and base radius $b_1 = 5$mm. The apex radius of curvature of the apex is $R_a^{(2)} = 0.02$mm.
The scaled field is plotted till $\rho = 1.25R_a^{(2)}$ since field emission is generally negligible
for $\rho > R_a$.

\begin{figure}[hbt]
  \begin{center}
    \vskip -3.0cm
\hspace*{-0.40cm}\includegraphics[scale=0.445,angle=0]{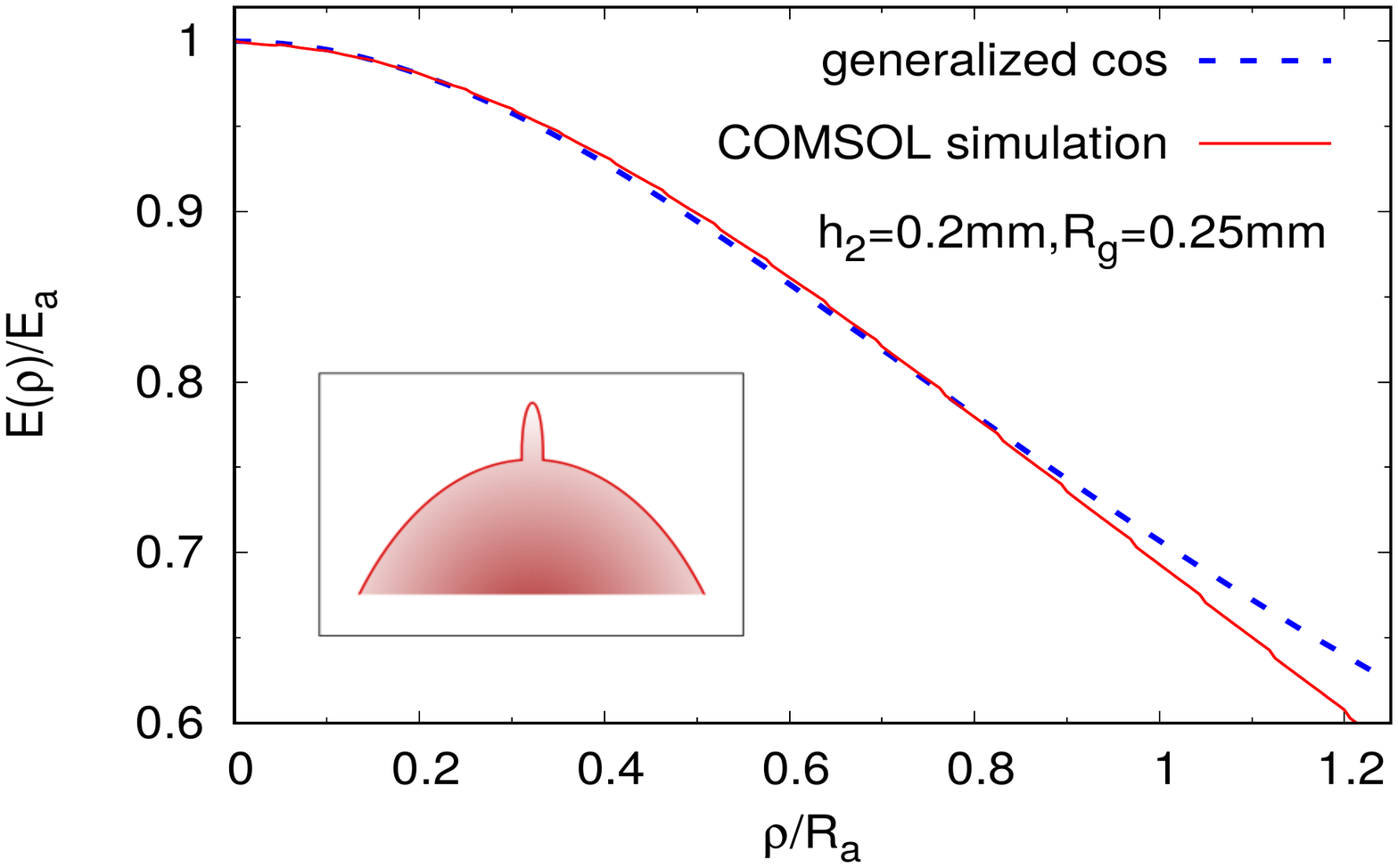}
\vskip -4.75cm
\caption{As in Fig.~\ref{fig:cos_hp0.5_Rg1} for $h_2 = 0.2$mm and $R_g = 0.25$mm.
  Deviation from the cosine law occurs for $\rho > 0.8 R_a$.
  The inset shows the hemiellipsoid protrusion and the endcap of the hemiellipsoid base. 
  }
\label{fig:cos_hp0.2_Rg0.25}
  \end{center}
\end{figure}

Clearly, the local field follows the generalized cosine law very well. A similar observation holds for other aperture
radius such as $R_g = 0.25$mm. Fig.~\ref{fig:cos_hp0.2_Rg0.25} shows the cosine law for a smaller
protrusion of height $h_2 = 0.2$mm at $R_g = 0.25$mm. The local field field follows the generalized cosine
law of Eq.~\ref{eq:cos} fairly well for $\rho < 0.8R_a$ but deviates thereafter. While this
is likely to induce only small errors in the field emission current for moderate fields,
the cause of the deviation from the cosine law is likely the effect of smaller gate aperture
and height of protrusion resulting in a
sharp change in line charge density near the top of the base.

\begin{figure}[h]
  \begin{center}
    \vskip -2.75cm
\hspace*{-0.50cm}\includegraphics[scale=0.44,angle=0]{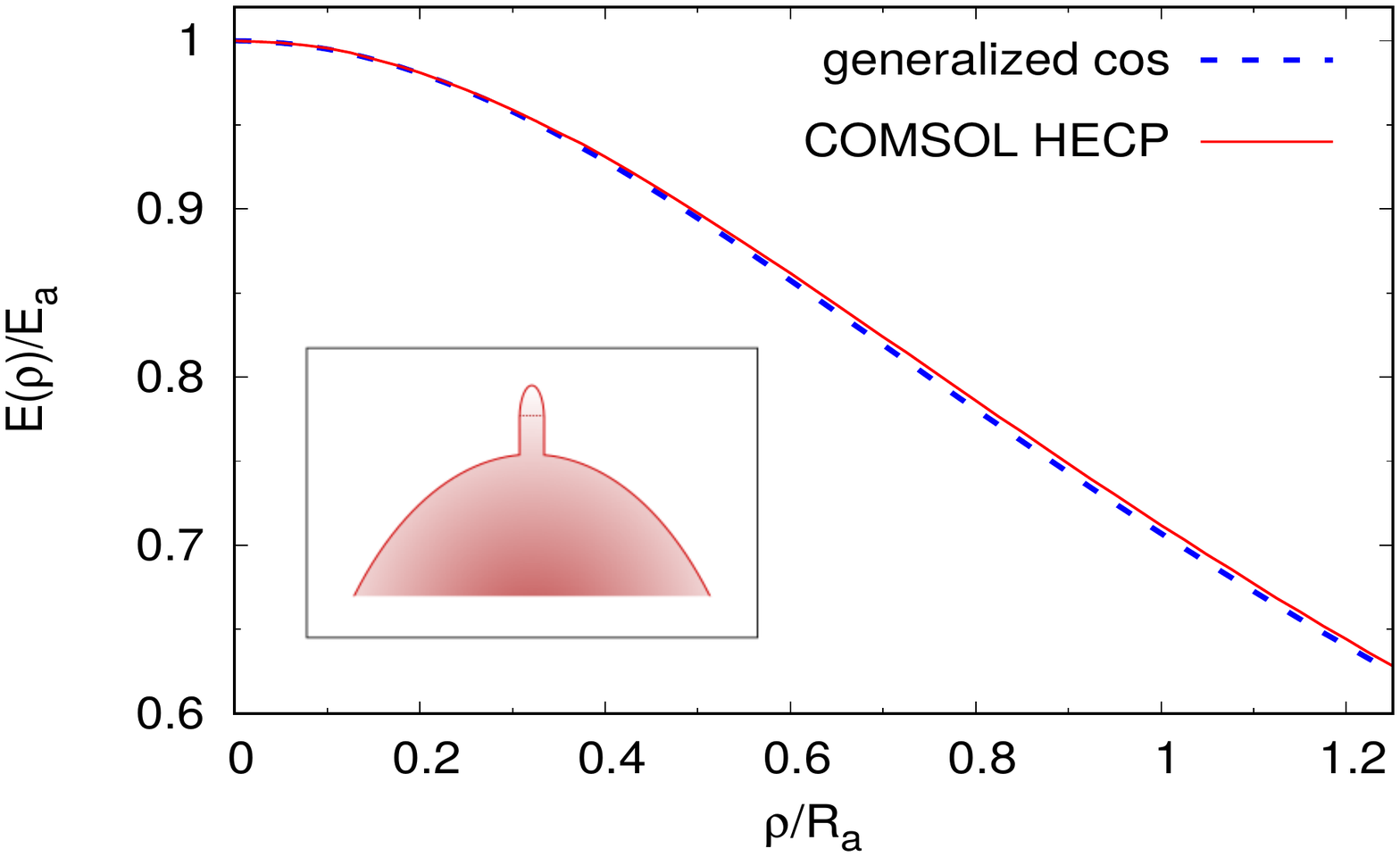}
\vskip -4.450cm
\caption{The variation in local field for an HECP protrusion for $h_2 = 0.5$mm comprising of
  a cylindrical post of height 0.3mm, radius = 0.05mm and a hemiellipsoid endcap of height
  0.2mm and base radius 0.05mm. Here $D = 51$mm and the gate radius $R_g = 1$mm.
  The inset shows the HECP protrusion and the endcap of the hemiellipsoid base.
  }
\label{fig:cos_hp0.5_Rg1_HECP0.5}
  \end{center}
\end{figure}

\vskip -0.5cm
We next consider an HECP protrusion on a hemiellipsoid base. The HECP has a total height $h_2 = 0.5$mm. The
cylindrical post has a radius 0.05mm and height 0.3mm. The ellipsoid cap on HECP thus has a height 0.2mm,
base radius 0.05mm and apex radius of curvature $12.5\mu$m. Thus, the hemiellipsoid endcap
has a height 8 times the apex radius of curvature. The variation of the local
field is shown in Fig.~\ref{fig:cos_hp0.5_Rg1_HECP0.5} (solid line) along with the generalized
cosine law. The agreement is good. As the height of the hemi-ellipsoid endcap becomes smaller
than about five times the apex radius of curvature and tends towards a hemispherical endcap,
deviations from the cosine law become larger.

\section{Field emission from a compound cathode and particle tracking}

It is thus clear that a compound emitter with a locally parabolic endcap follows the generalized
cosine law provided the endcap is not too close to the junction of two stages.
This can be used to determine the net emission current using the electric field and the radius
of curvature at the emitter apex. 
The net field emission current from the endcap using the local current density is given by

\be
I_{MG} = \int_0^{R_a} d\rho~J_{MG}(\rho)~ 2\pi \rho  \sqrt{1 + \left(\frac{dz}{d\rho}\right)^2}
\ee

\noi
where the Murphy-Good expression for the current density

\be
J_{MG}(\rho) = \frac{A_{FN}}{ t_{\small F}^2(\rho)} \frac{E^2(\rho)}{\phi} \exp(-{\it v}_{{\small F}}(\rho) B_{{\small FN}}\phi^{3/2}/E(\rho)) \label{eq:MG}
\ee

\noi
is used and the field variation follows the generalized cosine law

\be
E(\rho) = E_a \frac{z/h}{\sqrt{(z(\rho)/h)^2 + (\rho/R_a)^2}}. \label{eq:cos1}
\ee

\noi
Here, $A_\fn~\simeq~1.541434~{\rm \mu A~eV~V}^{-2}$ and 
$B_\fn \simeq 6.830890~{\rm eV}^{-3/2}~{\rm V~nm}^{-1}$ are the conventional Fowler-Nordheim constants,
while ${\it v}_{{\small F}}(\rho)  =  1 - f(\rho) + (f(\rho)/6)\ln f(\rho)$,
$t_{\small F}(\rho)  =  1 + f(\rho)/9 - (f(\rho)/18)\ln f(\rho)$ are corrections due to the image-charge potential
with $f(\rho)  =  1.439965 E(\rho)/\phi^2$.
We shall assume the workfunction $\phi = 4.5$eV hereafter.
Note that  in Eq.~(\ref{eq:cos1}), $z = h - \rho^2/(2R_a)$ for a locally parabolic tip.

For purposes of simulation, the net current must be generated by appropriate distributions
of emitted electrons from the surface. For sharp emitters, the
distribution\cite{DB_parabolic}

\be
f(\tth) d\tth  \simeq  2\pi R_a^2~\frac{\sin\tth}{\cos^2\tth} \frac{A_\fn}{\phi}\frac{E_a^2}{t_F^2(\tth)} e^{-\frac{B_\fn \nuv_F(\tth) \phi^{3/2}}{(E_a \cos\tth)}} d\tth \label{eq:angular}
\ee

\noi
can be used to determine the current

\be
\Delta I = \int_{\tth}^{\tth + \Delta \tth} f(\tth) d\tth
\ee

\noi
emitted between $\tth$ and $\tth + \Delta\tth$ from the surface of the emitter tip.
In Eq.~(\ref{eq:angular}), the free electron model is assumed,
barrier lowering due to the image potential is incorporated
and it is assumed that the surface near the apex has a uniform workfunction $\phi$.

Note that the $\rho$ and $\tth$ are related through Eq.~(\ref{eq:cos}) so that the
distribution can equivalently be expressed in terms of $\rho$.
For sharp parabolic emitters ($R_a/h << 1$), $\rho \simeq R_a \tan\tth$. Using this relation
between $\rho$ and $\tan\tth$, it follows that the angle $\theta_L$ that
the normal (at any point $\rho$ on the parabolic surface $z = h - \rho^2/2R_a$)
makes with the emitter axis, is such that $\tan\theta_L = \rho/R_a \simeq \tan\tth$.
Thus, $\theta_L \simeq \tth$ so that
$f(\tth)$ also describes the distribution of launch angles ($\theta_L$) of electrons from the
surface of a parabolic emitter\cite{DB_parabolic,corrections}.

Having generated a value of $\tth$ using the distribution in Eq.~(\ref{eq:angular}), a value
for the normal energy $\calE_N$ and total energy $\calE_T$ can be generated using the distributions
$f(\calE_N|\tth)$  and $f(\calE_T|\calE_N,\tth)$, both of which can be arrived at using the joint
distribution\cite{DB_parabolic} $f(\tth,\calE_N,\calE_T)$. Thus the velocity components
in a local co-ordinate system centred at ($\rho(\tth),z(\tth)$) on the surface
of the emitter can be found using $\tth$, $\calE_N$ and $\calE_T$.
For a given simulation time-step $\Delta T$, the net charge emitted from $\tth$ and $\tth + \Delta\tth$
is $\Delta Q = \Delta I \Delta t$. The charge $\Delta Q$ can be distributed among a
number of macro-particles with a distribution of velocities.

The algorithm described above has been incorporated in the
Particle-In-Cell code PASUPAT\cite{benchmark,open_diode,PoP_PASUPAT} which allows
emission of electrons (macroparticles with variable weight in general) from the
endcap of the structures
with a knowledge of only the apex electric field $E_a$. For a compound multi-scale geometry
such as the one considered here, the local field around the apex of is upto
three orders of magnitude higher than the macroscopic field ($\simeq V_g/D$).
Thus, an Yee-Grid based finite difference approach is not feasible
for the entire triode region, comprising of the cathode plate with the compound emitter, the
gate and the anode. This is especially so due to the enormous resources
required for accurate tracking of the particles in the immediate neighbourhood of the emitter apex.
Note that an error in the electrostatic field within about a distance $R_a$ from the apex
can adversely affect energy conservation and the emittance.
We shall therefore use a hybrid approach to study particle dynamics for
such multiscale cathodes.

\begin{figure}[hbt]
  \begin{center}
    \vskip -2.6cm
\hspace*{0.30cm}\includegraphics[scale=0.45,angle=0]{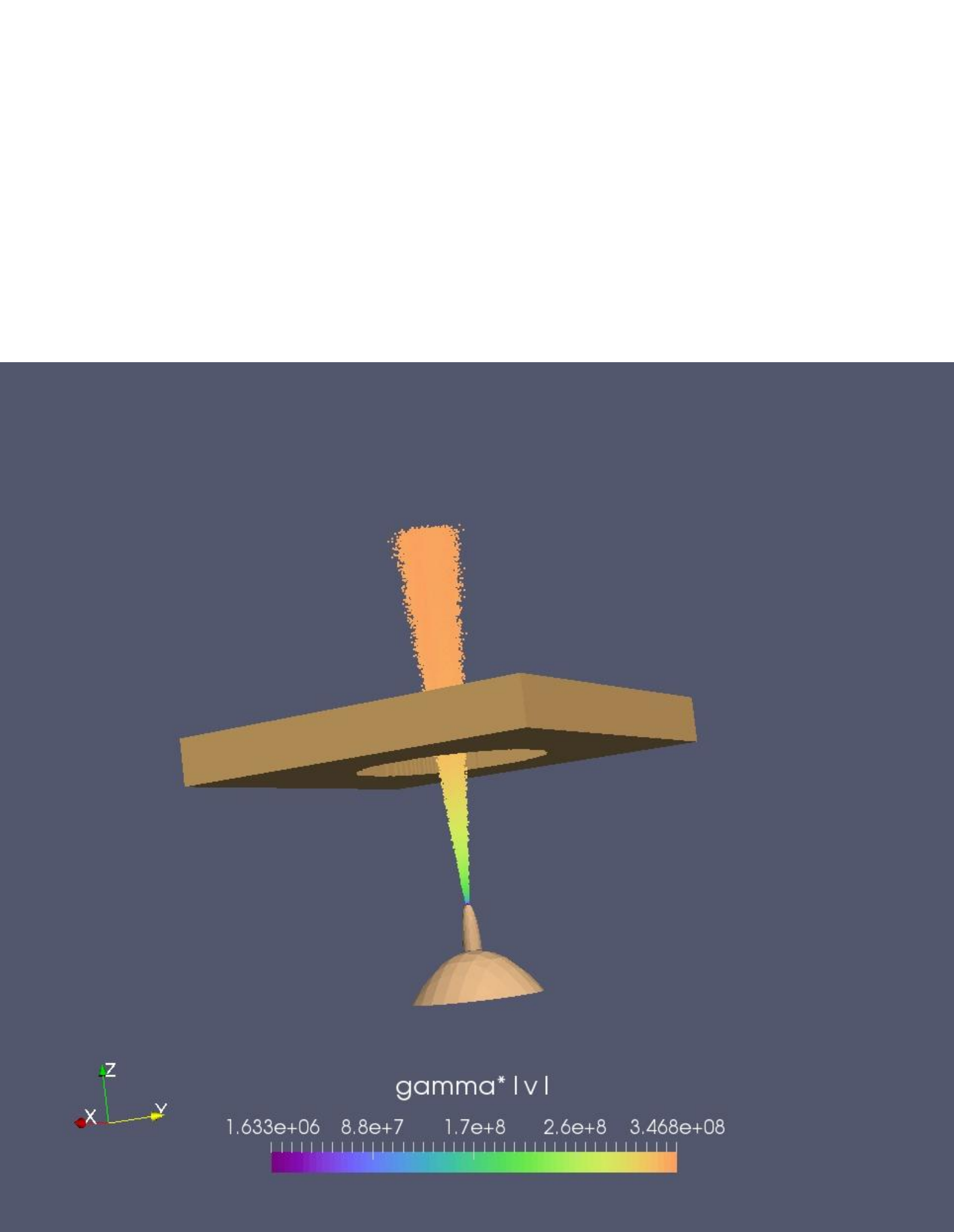}
\caption{A PIC simulation of a compound emitter using PASUPAT where macro-particles are
  emitted using distributions
  $f(\tth)$, $f(\calE_N|\tth)$ and $f(\calE_T|\tth,\calE_N)$. The hemiellipsoid protrusion
  is shown in full while only the top portion of the hemiellipsoid-base is visible.
  The parameters used are $D=52$mm, $h_2 = 0.5$mm,
  $b_2 = 0.1$mm, $h_1 = 50$mm, $b_1 = 5$mm and $R_g = 1$mm.
  At each time step 500 macroparticles are emitted. The anode and gate are grounded while the cathode
  is at a potential $V_g = -260$kV. The field at the apex of the protrusion is $E_a \simeq 3.79$V/nm.
  }
\label{fig:PIC}
  \end{center}
\end{figure}

In the approach adopted here, the applied electrostatic vacuum field for the
triode is evaluated separately using a finite element code. The electric
field components are then imported to
the PIC code PASUPAT where a truncated geometry is used with $z \geq z_T$.
Thus for the 2-stage multiscale cathode having a hemiellipsoid protrusion of
height $h_2 = 0.5$mm on a hemiellipsoid base of height $h_1 = 50$mm, the truncated
geometry simulated corresponds to $z \geq z_T$ with $z_T =  49.495$mm.
The geometry simulated (in PASUPAT) therefore consists of the
base-endcap, the protrusion, the gate and the anode. The mesh chosen is 
fine enough for accurate tracking but too coarse for field emission.
After particle emission begins, space charge effects are evaluated
separately by solving Poisson equation with
all conductors grounded and `open electrostatic boundary' condition\cite{ABC,pop2015} imposed on the plane
$z = z_T$ excluding the portion occupied by the base-endcap which is a conductor. The space charge field is
then added to the applied field and used for subsequent emission at the next
time-step and for particle tracking.
Fig.~\ref{fig:PIC} shows a typical beam-optics simulation using PASUPAT where particles are
emitted and tracked using the methodology described. The colour scale shows the
speed multiplied by the relativistic factor `gamma'.
In this case, space charge effects do not significantly affect the emission current
and only mildly affects the beam as it passes through the gate aperture ($R_g = 1$mm).

\begin{figure}[hbt]
  \begin{center}
    \vskip -0.75cm
\hspace*{-0.70cm}\includegraphics[scale=0.35,angle=0]{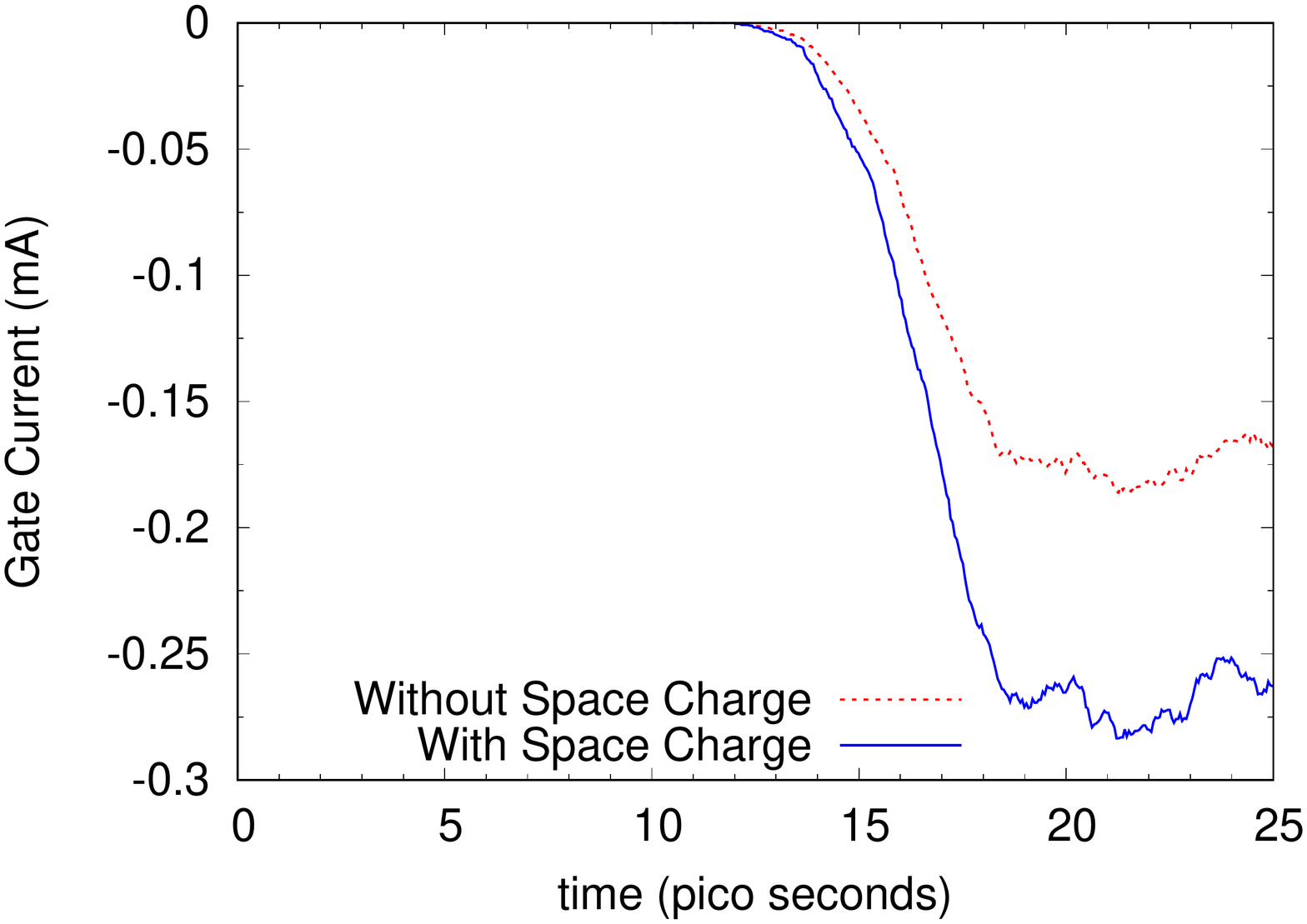}
\vskip -0.50cm
\caption{The gate current for gate radius $R_g = 0.25$mm and $D = 52$mm, with and without
  space charge effect. The gate loss increases when space charge effect is turned on.
  The field at the apex of the protrusion is $E_a \simeq 3.85$V/nm.
  }
\label{fig:gate_loss}
  \end{center}
\end{figure}

As the gate aperture is reduced, the vacuum apex field increases. While the space charge
field at the apex is not strong enough to affect the emitted current significantly, the gate
current loss increases when space charge is switched on. This can be seen in Fig.~\ref{fig:gate_loss}
where the locally time-averaged gate current (the current lost at the gate) with and without
space charge effects is shown for $D = 52$mm with $R_g = 0.25$mm.

\begin{figure}[hbt]
  \begin{center}
    \vskip -0.75cm
\hspace*{-0.70cm}\includegraphics[scale=0.35,angle=0]{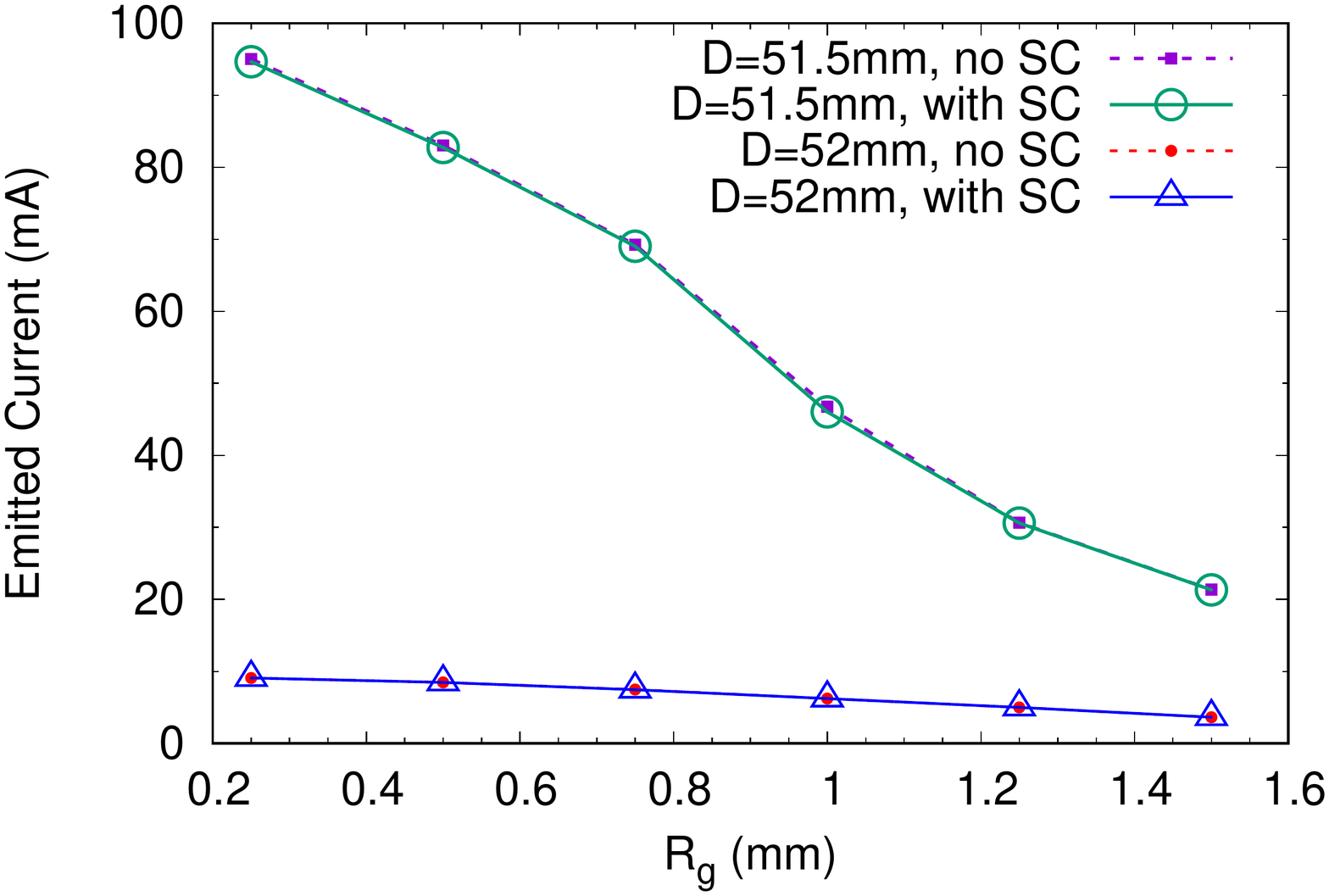}
\vskip -1.0cm
\caption{The emitted current as a function of gate radius $R_g$ for
  two gate locations. The total height of the compound emitter is 50.5mm with the
  hemiellipsoid protrusion having a height $h_2 = 0.5$mm and base radius $b_2 = 0.1$mm.
  The cathode potential $V_g = -E_0 D$ where $E_0 = 5 \times 10^6$V/m. Space charge has
  very little effect on emitted current. For comparison, at $R_g = 0.25$mm, $E_a \simeq 3.85$V/nm
  for $D = 52$mm while for $D = 51.5$mm, $E_a \simeq 4.41$V/nm.
  }
\label{fig:PASUPAT_current}
  \end{center}
\end{figure}

Finally, Fig. \ref{fig:PASUPAT_current}  shows the emitted current for different
gate aperture radius $R_g$ for a hemi-ellipsoid
protrusion with $h_2 = 0.5$mm and $b_2 = 0.1$mm on a hemiellipsoid base with $h_1 = 50$mm
and $b_1 = 5$mm. The compound emitter and cathode plates are kept at a
potential $V_g = -E_0 D$ where $E_0 = 5\times 10^6$V/m. 
The variation in emitted current as the gate aperture $R_g$ is varied is shown for 2 different
gate locations $D = 51.5$mm and $D = 52$mm.
Clearly, the emitted current decreases with an increase in gate aperture and
distance from the tip of the apex of the compound structure. Note that space
charge effects do not significantly alter the emitted current for $D = 52$mm and
has only a mild effect for $D = 51.5$mm for small gate apertures.

\section{Discussions and Conclusions}

We have demonstrated a methodology to simulate field emission from compound multi-stage
emitters using a hybrid approach (such as in Ref [\onlinecite{db_hybrid}]) wherein
a combination of analytical and numerical methods can be used to emit and track
particles in the gated diode region. We have further adapted this approach to
include space charge effects using a truncated geometry with `open electrostatic
boundary condition'.

Crucial to this task was the demonstration that the generalized cosine law of
local field variation on the surface of locally parabolic tips, continues to hold in multistage emitters.
This enables us to use the results for the net current distributions of electrons in position
and velocity and emit macroparticles from the protrusion where the local field is high.
Note that the requirement of a locally parabolic tip excludes protrusions with
hemi-spherical endcaps but is otherwise generic. Also, the base of the axially symmetric
multiscale emitter can assume shapes other than the hemiellipsoid considered here.
Our numerical investigations also show that the Corrected Schottky Conjecture performs well in predicting
the apex field enhancement factor for compound emitters in a gated diode and can be used
for predicting field emission currents with reasonable accuracy.

While the simulation methodology for multiscale emitters developed here is sufficiently general, we
have restricted ourselves to problems where space charge does not alter the apex field substantially.
The use of distributions to emit electrons requires the generalized cosine law to hold in the
presence of space charge arising from field emission. It thus needs to be independently established that the
cosine law holds at moderate to high vacuum field strengths where the current emitted is much
larger.

\vskip 0.25 in

\noi
    {\it Author Credits} \textemdash~ 
    S.S carried out the COMSOL simulations on corrected Schottky conjecture and generalized cosine law; R.K. and G.S. adapted PASUPAT
    for multi-stage cathodes and performed the simulations;
    D.B. conceived the idea, devised the methodology, supervised the work; all authors contributed to the
    manuscript. \\ 

\vskip 0.25 in

\noi
    {\it Acknowledgements} \textemdash~
    PASUPAT simulations were performed on ANUPAM-AGANYA super-computing facility at Computer Division, BARC.
    
\vskip 0.25 in
\noi
{\it Data Availability}: The computational data that supports the findings of this study are available within the article.

\section{References}
\vskip -0.25 in


\begin{thebibliography}{99}

\bibitem{Schottky23} W. Schottky, Z. Phys. 14, 63 (1923).
\bibitem{stern} T.~E.~Stern, B.~S.~Gossling and R.~H.~Fowler, Proc. R. Soc. Lond. A 124, 699 (1929).
\bibitem{miller09} R.~Miller, Y.~Y.~Lau, and J.~H.~Booske, J. Appl. Phys. 106, 104903 (2009).

\bibitem{deAssis16} T.~A. de Assis and F.~F.~Dall'Agnol, Nanotechnology 27, 44LT01 (2016).
\bibitem{marcelino} E.~Marcelino, T.~A.~de Assis, and C.~M.~C.~de~Castilho, J. Vac. Sci. Technol. B 35, 051801 (2017).
\bibitem{harris19a} J.~R.~Harris and K.~L.~Jensen, Journal of Applied Physics 125, 215306 (2019).
\bibitem{zhu2019} W. Zhu, M. Cahay, J. Ludwick, K.L. Jensen, R.G. Forbes, S.B. Fairchild, T.C. Back, P.T. Murray, J.R. Harris and D.A. Shiffler, ``Multiscale modeling of field emission properties of carbon-nanotube-based fibers'' in {\it Nanotube Superfiber Materials}, 2nd Edition, William Andrew Publishing, pp. 541-572 (2019).
\bibitem{marcelino19} E.~Marcelino de Carvalho Neto, Journal of Applied Physics 126, 244502 (2019)  
\bibitem{DB_CSC_20} D.~Biswas, J. Vac. Sci. Technol. B 38, 023208 (2020).
\bibitem{CT_2020} S.~Sarkar, R.~Kar, J.~Mondal, L.~Mishra, D.~Jayaprakash, N.~Maiti, R.~Tripathi, D.~Biswas, Carbon Trends (in press) https://doi.org/10.1016/j.cartre.2020.100008.
\bibitem{jensen2019a} K.~L.~Jensen, M.~McDonald, J.~R.~Harris, D.~A.~Shiffler, M.~Cahay, J.~J.~Petillo, J. Appl. Phys. 126, 245301 (2019).
\bibitem{ultram} D.Biswas, G.Singh, S.Sarkar and R.Kumar, Ultramicroscopy 185,1-4 (2018).
\bibitem{physE} D.~Biswas, G.~Singh and R.~Ramachandran, Physica E, 109, pp. 179 (2019).
\bibitem{FN28} R. H. Fowler, L. Nordheim, Proc. R. Soc. A 119,  173 (1928).
\bibitem{MG56} E. L. Murphy, R. H. Good, Phys. Rev. 102,  1464 (1956).
\bibitem{RF2006} R.~G.~Forbes,  App. Phys. Lett. 89, 113122 (2006).
\bibitem{RFD2007} R.~G.~Forbes and J.~H.~B.~Deane, Proc. Roy. Soc. A 463, 2907 (2007).
\bibitem{KLJ2014} K.~L.~Jensen, {\it Field emission - fundamental theory to usage}, Wiley Encycl. Electr. Electron. Eng. (2014).
\bibitem{DBRR2019} D. Biswas, R. Ramachandran, J. Vac. Sci. Technol. B 37, 021801 (2019).
\bibitem{RF2019} R. Forbes, J. Appl. Phys,126 (21) 210901  (2019).
\bibitem{DB_parabolic} D. Biswas, Phys. Plasmas, 25, 043105 (2018).
\bibitem{egorov17} N.~V.~Egorov, A.~Y.~Antonov, and N.~S.~Demchenko, Tech. Phys. 62, 201 (2017).
\bibitem{DB_universal} D.Biswas, Physics of Plasmas 25, 043113 (2018).
\bibitem{DBSG2019} D.~Biswas and S.~Sarkar, J. Vac. Sci. Technol. B 37, 062203 (2019) 
\bibitem{hemiellip} For a hemiellipsoid emitter, $\alpha_1 = 1$ and $\alpha_2 = 2$ when $h/R_a \gtrapprox 50$.
\bibitem{csc_averaging} The Corrected Schottky Conjecture may perform better for certain gated
  structures when the averaging interval is reduced, for instance from $h_2$ to $3h_2/4$.
\bibitem{db_anode} D.Biswas, Phys. Plasmas 26, 073106 (2019). 
\bibitem{db_jap2016} D.Biswas, G.Singh and R.Kumar, J.~Appl.~Phys (2016). 
\bibitem{zheng2020} F.~Zheng, G.~Pozzi, V.~Migunov, L.~Pirker, M.~Rem\v{s}kar, M.~Beleggia and R.~E.~Dunin-Borkowskia, Nanoscale, 12, 10559 (2020).
\bibitem{surface2line} The line charge density can be arrived at by projecting the surface charge density which
  in turn can be determined from the normal electric field on the surface of the emitter in a numerical calculation.
  For an example see Ref. [\onlinecite{db_jap2016}]. A nonlinear line charge density has been
  observed experimentally in Ref. [\onlinecite{zheng2020}].
\bibitem{corrections} Corrections to the distribution $f(\tth)$ and the launch angle
  need to be incorporated for emitters that are not sharp. Eq.~(\ref{eq:angular}) is adequate
  for emitters with apex field enhancement factors greater than fifty.
\bibitem{benchmark} PASUPAT has been benchmarked using standard problems. It reproduces the
  Child-Langmuir law for planar diode, both for the finite and infinite emission area and
  correctly predicts the electromagnetic power loss from open diodes \cite{open_diode}.
  It has recently been used to verify the linear scaling of space charge limited current
  with the apex field enhancement factor \cite{PoP_PASUPAT}.
\bibitem{open_diode} D.~Biswas, R.~Kumar and R.~R.~Puri, Physics of Plasmas 12, 093102 (2005).
\bibitem{PoP_PASUPAT} G.~Singh, R.~Kumar and D.~Biswas, Physics of Plasmas 27, 104501 (2020).
\bibitem{ABC} A.~Khebir, A.~B.~Kouki, and R.~Mittra, IEEE Trans. Microwave Theory Tech. 38, 1427 (1990).
\bibitem{pop2015} D.~Biswas, G.~Singh and R.~Kumar, Phys. Plasmas 22, 093119 (2015) 
\bibitem{db_hybrid} D.~Biswas, J. Vac. Sci. Technol. B 38, 063201 (2020).
\end{thebibliography}
\end{document}